\documentclass[reprint,prl,aps,epsf,twocolumn,superscriptaddress]{revtex4-1}

\usepackage{graphicx}
\usepackage{amssymb}

\newcommand{\omi}{\mbox{\boldmath $\Omega_i$}}

\newcommand{\HT}{\mbox{\boldmath $H$}}
\newcommand{\QTa}{\mbox{\boldmath $Q_1$}}
\newcommand{\QTb}{\mbox{\boldmath $Q_2$}}

\begin{document}
\title{Kinetic Theory of Flocking: Derivation of Hydrodynamic Equations}
\author{Thomas Ihle}
\affiliation{Department of Physics, North Dakota State University,  
Fargo, North Dakota, 58108-6050, USA}    
\affiliation{Max Planck Institute for the Physics of Complex Systems, N{\"o}thnitzer Stra{\ss}e 38,
01187 Dresden, Germany}  

\begin{abstract}
It is shown how to explicitly coarse-grain the microscopic dynamics of the rule-based Vicsek model for self-propelled agents.
The 
hydrodynamic equations 
are 
derived
by means of an Enskog-type kinetic theory.
Expressions for all transport coefficients at large particle speed 
are given.
The 
phase transition from a disordered to a flocking state is studied
numerically and analytically. 
\end{abstract}

\maketitle
Pattern formation and 
collective motion 
in systems of self-propelled 
objects 
are fascinating phenomena
which have attracted much attention. 
Systems of interest include 
animal flocks \cite{animal_flocks},  
chemically powered nanorods \cite{nano_rods}, and 
actin networks driven by molecular motors \cite{actin_net}. 
Theoretical studies of these systems
are usually based on  
phenomenological transport equations. 
In most cases, the equations are postulated by means of symmetry 
arguments, which
define only the general form of the terms 
but
leave their coefficients undetermined.

One goal of this Letter is to provide a systematic derivation of 
all relevant coefficients 
for the two-dimensional Vicsek model (VM) of self-propelled particles \cite{vicsek_95_07}.
In the VM, pointlike particles are driven with  
constant speed. At each time step, a given particle assumes the average 
direction of motion of its neighboring particles, 
with some added noise.
As the noise amplitude decreases, the system undergoes a phase transition from a 
disordered state, in which the particles have no prefered global direction, to an
ordered state, in which
the particles move collectively in the same direction. 
This
long-range order  
motivated renormalization group studies
by Toner and Tu \cite{toner_95_98}.
They found that the stabilization of the ordered phase
is due to the nonzero speed of the particles, 
allowing two originally distant particles to interact with each other at a later time.
The phase transition was originally thought to be continuous \cite{vicsek_95_07} but
recent numerical work \cite{chate_04_08} indicates that the transition is discontinuous
with strong finite size effects. 
There are few
analytical studies on this transition \cite{bertin_06,aldana_07}. 
They do not treat the original VM but simple models related to it.
For example, Bertin {\em et al.} \cite{bertin_06}, study a model with 
simplified interactions and
a continuous time dynamics  
by means of a Boltzmann equation.

Numerical simulations of the VM \cite{vicsek_95_07, chate_04_08} 
show localized high-density structures,
for which a Boltzmann description, which is restricted to low densities, is not sufficient.
Enskog's proposal to generalize the Boltzmann equation to dense gases 
was a major milestone in kinetic theory.
In this Letter, it is shown 
how
an
Enskog-type equation with genuine {\it multi-body collisions} can be obtained for the VM and  
how this can be used to
rigorously derive hydrodynamic equations. 
In addition to the terms postulated by Toner and Tu \cite{toner_95_98}, the 
derived
equations contain several new relevant  
terms which 
describe
an intricate coupling between density and order parameter gradients.
The coefficients of {\it all} 
terms, 
compatible with the symmetries of the system,
are calculated 
explicitly in third order of a gradient expansion. 
The new kinetic equation is used to determine the 
mean-field phase diagram of the 
VM, 
which agrees 
well with direct numerical
simulations but disagrees with
the results of a related continuous time model \cite{bertin_06}.  
This shows
the importance of explicitely taking the {\it discrete time, rule-based} nature
of the VM into account.
The derived hydrodynamic equations are applied to study the stability
of a homogeneous flocking state against spatio-temporal perturbations.
I discuss how an instability 
at the onset of collective motion can change the appearance
of the phase transition from second to first order. 
Predictions for the system size where this change is expected to happen,
are given.

In the VM, a 
system of $N$ pointlike particles with continuous 
spatial coordinates ${\bf r}_i(t)$ and velocities ${\bf v}_i(t)$ evolves via two
steps:
streaming and collision. 
During a time step $\tau$, particles stream ballistically:
${\bf x}_i(t+\tau)={\bf x}_i(t)+\tau {\bf v}_i(t)$.
The magnitude of the particle velocities is fixed to $v_0$. 
Only the directions $\theta_i$ of 
the velocity vectors 
are updated in the collision step:
a circle of radius $R$ is drawn around a given particle and the average direction $\bar{\theta}_i$ of motion
of the 
particles 
within the circle is determined
according to 
$\bar{\theta}_i={\rm arctan}[\sum_j^n {\rm sin}(\theta_j)/\sum_j^n {\rm cos}(\theta_j)]$. 
The new directions 
follow as
$\theta_i(t+\tau)=\bar{\theta}_i(t)+\xi_i$, where $\xi_i$ is a random number
chosen with uniform probability from the interval $[-\eta/2,\eta/2]$.
Since explicitly coarse-graining the dynamics of the VM
is 
difficult,
in previous work \cite{ihle_09},
I have first validated the 
formalism on a simpler equilibrium model 
\cite{mpc_sum}
which shares essential features with the VM.
The kinetic formalism 
starts with the 
Liouville equation for the N-particle probability density 

\begin{eqnarray}
\nonumber
& &P(\theta^{(N)}, {\bf X}^{(N)}+\tau {\bf V}^{(N)},t+\tau)= 
{1\over \eta^N}
\int_{-\eta/2}^{\eta/2}
d\xi^{(N)} \\
& & \times \int_0^{2\pi} d\tilde{\theta}^{(N)}
\,P(\tilde{\theta}^{(N)}, {\bf X}^{(N)},t) 
\label{PRL1}
\prod_{i=1}^N \hat{\delta}(\theta_i-\xi_i-\bar{\theta}_i) 
\end{eqnarray}
where 
${\bf X}^{(N)}\equiv({\bf x}_1,{\bf x}_2,\ldots, {\bf x}_N)$,
$\theta^{(N)}\equiv(\theta_1,\theta_2,\ldots, \theta_N)$, and
$\hat{\delta}(x)=\sum_{m=-\infty}^{\infty}\delta(x+2\pi m)$
is the periodically continued delta function. 
The velocities
${\bf V}^{(N)}\equiv({\bf v}_1,{\bf v}_2,..., {\bf v}_N)$, are given in terms
of angle variables, ${\bf v}_i=v_0(\cos{\theta_i}, \sin{\theta_i})$.
The collision integral contains integrations over
the pre-collisional angles $\tilde{\theta}_j$.
Assuming that the particles are uncorrelated 
{\it prior} to the collisions,
the probability distribution can be expressed as a product of identical one-particle 
probability distributions:
$P(\theta^{(N)}, {\bf X}^{(N)})
=\prod_{i=1}^N P_1(\theta_i, {\bf x}_i)$.
This approximation of {\em molecular chaos} 
is valid at moderate and large noise strength $\eta$ and 
when the mean free path (mfp)
is large 
compared to the radius of interaction $R$. 
Here, the mfp is defined as the distance a particle travels between collisions,
$\tau\,v_0$, and is {\em density-independent} due to the discrete nature of
the dynamics.
Multiplying Eq.\ (\ref{PRL1}) by 
$\sum_i\delta({\bf v}-{\bf v}_i)\delta({\bf x}-{\bf x}_i)$ and integrating 
over all particle positions $x_i$ and angles $\theta_i$, yields in the large $N$-limit 
\cite{ihle_09}, a kinetic equation for
the one-particle distribution function, 
$f(\theta,{\bf x},t)=NP_1(\theta,{\bf x},t)$, 
\begin{eqnarray}
\nonumber
& &f(\theta, {\bf x}+\tau {\bf v},t+\tau)=
{1\over \eta}
\int_{-\eta/2}^{\eta/2}
d\xi
\bigg\langle \bigg\langle
\sum_{n=1}^N
{{\rm e}^{-M_R}\over n!}
\,n\, \\
\label{ENSKOG1}
& &\times f(\tilde{\theta}_1,{\bf x}, t)
\,\hat{\delta}(\theta-\xi-\bar{\theta}_i)
\,\prod_{i=2}^n f( \tilde{\theta}_i, {\bf x}_i,t)
\bigg\rangle_{\tilde{\theta}} \bigg\rangle_x
\end{eqnarray}
where 
$M_R({\bf x},t)=\int_R\rho({\bf y},t)\,d{\bf y}$
is the average number of particles in a circle of radius $R$ centered around ${\bf x}$.
The local particle density $\rho$ is given as a moment of the distribution function,
$\rho({\bf x},t)=\int_0^{2\pi} f(\theta,{\bf x},t)\,d\theta$;
$\langle ... \rangle_x=\int_R... \,d{\bf x}_2\,d{\bf x}_3...d{\bf x}_n$
denotes the integration over all positions, $n-1$ particles can assume within the interaction circle;
$\langle ... \rangle_{\tilde{\theta}}=\int_0^{2\pi} ... d \tilde{\theta}_1 d \tilde{\theta}_2 ... 
d \tilde{\theta}_n $ 
is the average over all pre-collisional angles 
of $n$ particles in the interaction circle.
Since particles in the VM have zero volume, 
there is a non-zero probability that a large number of particles
can be found
in the collision circle of a given particle.
This leads to the unusual structure of the collision integral 
in which every term in the sum accounts for
a n-particle collision. For example, the $n=4$ term involves the product of four distribution functions
and describes a four body collision. 
Interactions between particles which are not at the same position but a distance $\le R$ apart
are explicitely taken into account by 
Eq.\ (\ref{ENSKOG1}).
This leads to collisional momentum transfer which is a key feature of the Enskog equation and not
included in Boltzmann-type equations.
Hence,
Eq.\ (\ref{ENSKOG1}), can be
interpreted as an 
Enskog-like equation for pointlike particles with discrete time evolution;
it remains valid
even at infinite density.

Let us first consider a spatially homogeneous system and study stationary 
solutions of Eq.\ (\ref{ENSKOG1}).
This amounts to solving the fixed-point equation $f_0(\theta)=C(f_0)$ for
the stationary distribution function $f_0$, where $C$ denotes 
the r.h.s. of Eq.\ (\ref{ENSKOG1}).
It can be easily checked that the constant distribution $f_0=\rho_0/2\pi$
is a fixed-point at any noise and average density,
$\rho_0=N/A$, where $A$
is the area of the system.
This solution corresponds to the disordered
phase, where all velocity directions occur at equal probability.
Below a critical 
noise $\eta_C(\rho_0)$
there exists another fixed-point solution which 
breaks rotational symmetry.
It has a maximum at some arbitrary angle $\hat{\theta}$ and describes
ordered motion into this direction.
The critical noise 
follows from the 
condition $\lambda=1$,
with
\begin{eqnarray}
\nonumber           
\lambda&=&{4 \over \eta}{\rm sin}\left({\eta\over 2}\right)
{\rm e}^{-M_R} \sum_{n=1}^N {n^2 M_R^{n-1}\over n!}\,I(n) \\
\label{LAMBDA_DEF}
I(n)&=&{1\over (2\pi)^n}    
\int_0^{2\pi}d\theta_1\ldots
\int_0^{2\pi}d\theta_n
\,\cos{\bar{\theta}}\,
\cos{\theta_1} 
\end{eqnarray}
Here, $M_R$ is equal to $\pi R^2 \rho_0$ and
$\bar{\theta}$ is the average angle defined above Eq. (\ref{PRL1}).
The fixed-point equation was solved numerically for $\eta\le \eta_C$.
The solution 
approaches a cosine with vanishing
amplitude when $\eta$ approaches the critical noise.
By means of a 
Fourier cosine series in $\theta-\hat{\theta}$ the behavior at the critical point was extracted analytically.
The order parameter, defined as the amplitude $g_1$
of the first non-trivial Fourier coefficient, is found to behave as
$g_1\propto \sqrt{\eta_C-\eta}$.
Thus, the order-disorder transition appears to be continuous with the 
mean-field critical exponent of $1/2$.
Fig.\ 1a) shows the calculated phase diagram (solid line).
Evaluating Eq. (\ref{LAMBDA_DEF}) in the low density limit
gives $\eta_C\propto R\sqrt{\rho}$.
This scaling with the square root of the density agrees with previous numerical 
\cite{vicsek_95_07} 
and theoretical results \cite{chate_04_08,bertin_06}.
However, there is no dependence of the critical noise on the particle speed 
in the {\it large mfp limit},
which 
is consistent with numerical simulations of the VM \cite{vicsek_95_07,ihle_10} but
disagrees with the scaling 
$\eta_C\propto \sqrt{\rho R v_0}$ 
for $\rho\rightarrow 0$
of the continuous model of Ref. \cite{bertin_06}.
The dashed line in Fig 1a) shows that the phase diagram of
this model (obtained from Eq. (35) in \cite{bertin_06} with $v_0\tau/R=5$) does
not describe the VM. 
Evaluating Eq. (\ref{LAMBDA_DEF})
in the infinite density limit  
yields
$\eta_C\rightarrow 2\pi$.  
\begin{figure}
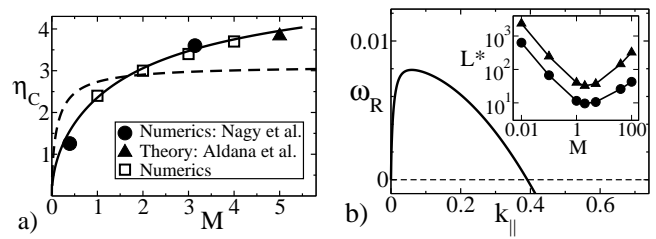

\begin{center}
\vspace{0cm}
\includegraphics[width=1.6in,angle=0]{PRL1aNew1.eps}
\hspace{1ex}
\includegraphics[width=1.6in,angle=0]{PRL1bnew.eps}
\vspace{-1ex}
\caption{a) The critical noise $\eta_C$ as a function of the average
number of collision partners, $M=\rho_0\pi R^2$, 
and the prediction of Eq. (35) for large $v_0$ from Ref. \cite{bertin_06}, 
(dashed line),
in comparison with results from Refs.\ \cite{vicsek_95_07, aldana_07, ihle_10}.
b) Real part of the growth rate, $\omega_R$, of a small longitudinal 
perturbation of the {\em ordered state}
versus dimensionless wave number $k_{||}$ at $M=5$,
very close to the threshold, 
$(\eta_C-\eta)/\eta_C=0.00057$.
The insert shows a lower and an upper bound for the crossover length
$L^{*}$ (in units of the mfp) beyond which the phase transition
is expected to become discontinuous.
}
\label{fig1}
\end{center}
\vspace*{-5ex}
\end{figure}
In order to see 
whether the homogeneous ordered state is stable
under time evolution, I derive the hydrodynamic equations by means of a Chapman-Enskog expansion \cite{mcquarrie_76,ihle_09}. 
The basic idea behind this expansion is to take the local stationary state as a reference state and expand
around it in powers of the hydrodynamic gradients.
To systematically account for these gradients 
a dimensionless ordering parameter $\epsilon$ is introduced, which is set to unity at the end
of the calculation.
The procedure  
starts with 
a Taylor expansion of the l.h.s of Eq. (\ref{ENSKOG1})
around $(\theta, {\bf x}, t)$.
The spatial gradients that occur are scaled as
$\partial_{\alpha}\rightarrow \epsilon\partial_{\alpha}$, 
and
multiple time scales $t_i$ are introduced in the temporal gradients.
These time scales describe different physical processes, for example, in regular fluids,
the time scale proportional to $\epsilon$ describes convection.
For the VM, this is expressed as
$\partial_t=\partial_{t_0}+\epsilon\partial_{t_1}+\epsilon^2\partial_{t_2} \ldots\, .
$

Expanding the distribution function and the collision integral in powers 
of $\epsilon$, $f=f_0+\epsilon f_1 +\epsilon^2 f_2+...$, 
and $C=C_0+\epsilon C_1+\epsilon^2 C_2+...$, inserting into 
Eq.\ (\ref{ENSKOG1}), and collecting terms of the same order
in $\epsilon$ leads to a hierarchy of evolution equations for the $f_i$.
Due to the absence of momentum conservation and Galilean invariance this
set 
of equations is dramatically different from the usual one. 
It is not {\em a priori} evident whether the scaling ansatz 
for the time derivatives is
correct.
However, it turns out that this choice 
avoids any inconsistencies 
if additionally 
the expansion of the distribution function $f$ is
identified as an angular Fourier series
with 
$f_0({\bf x},t)=\rho({\bf x},t)/ 2\pi$ and, for $n>0$,
$f_n({\bf x},\theta,t)=\left[a_n({\bf x},t)\cos{(n\theta)}+
b_n({\bf x},t)\sin{(n\theta)}\right]/\pi v_0^n
$.

Many moments of the collision integral such
as $\langle v_x v_y C_2\rangle=\int_0^{2\pi}v_x v_y C_2\,d\theta$ 
are required in the Chapman-Enskog expansion. 
For simplicity, these moments are evaluated in the {\it limit of large mfp}, $\tau\,v_0\gg R$.
This involves solving the following four 
integrals, 
\begin{equation}
\label{INTEGRALS}
J_m(n)={1\over (2\pi)^n}    
\int_0^{2\pi}d\theta_1\ldots
\int_0^{2\pi}d\theta_n\,
\Psi_m
\end{equation}
where $\Psi_m$ is given
by $\Psi_1=
\cos^2{\bar{\theta}}\,
\cos{2\theta_1}$, 
$\Psi_2=
\cos{\bar{\theta}}\,\sin{\bar{\theta}}\,
\cos{\theta_1}\,\sin{\theta_2} 
$,
$\Psi_3=
\cos{\bar{\theta}}\,
\cos{\theta_1}\,\cos{2 \theta_2} 
$, and
$\Psi_4=
\cos{\bar{\theta}}\,
\cos{\theta_1}\,\cos{\theta_2}\,\cos{\theta_3}$.
The average angle
$\bar{\theta}$ is a function of the angles
$\theta_1,\theta_2,\ldots \theta_n$.

We seek a hydrodynamic description of the first two moments of $f$, namely the particle density
$\rho=\int_0^{2\pi}f\,d\theta $
and the macroscopic momentum density vector $\vec{w}=(w_x,w_y)$,
$\vec{w}=\int_0^{2\pi}\vec{v} f\,d\theta. $
Inserting the Fourier representation of $f$ into these moments shows that 
the first order coefficients are given by the momentum density, $a_1=w_x$ and $b_1=w_y$. 
Multiplying the hierarchy of evolution equations by powers of the microscopic velocity
vector $\vec{v}=(v_x, v_y)$ and integrating over $\theta$ gives a set of equations
for the time development of the density and the moments $a_i$ and $b_i$.
This analysis is performed
in the vicinity of the critical point, $|\lambda-1|\ll 1$, in order to
significantly simplify 
the consistent closure of the hierarchy of moment equations, see \cite{ihle_10}.

For simplicity, all equations are rescaled by expressing time 
in units of $\tau$ and distances in units of the mfp, $\tau v_0$,
which also makes $\rho$ and $\vec{w}$ dimensionless.
After straightforward, but tedious, calculations one obtains
the continuity equation $\partial_t\rho+\partial_{\alpha}
w_{\alpha}=0$, and a rotationally-invariant equation for the 
momentum density,
\begin{equation}
\label{FINAL_EQ}
\partial_{t}\vec{w}+\nabla\cdot\HT=-b\nabla\rho+(\lambda-1)\vec{w}+
\QTa\cdot \vec{w}+\QTb\cdot\nabla\rho
\end{equation}
with $b=(3-\lambda)/4$.
The momentum flux tensor $\HT$ and the tensors $\QTa$, $\QTb$,
\begin{equation}
\label{TENSOR1}
\HT=\sum_{i=1}^5h_i\,\omi\;\;\;\;\;\;\;
\QTa=\sum_{i=1}^5q_i\,\omi \;\;\;\;\;\; 
\QTb=\sum_{i=1}^5k_i\,\omi
\end{equation}
are given in terms of five symmetric traceless tensors $\omi$, 
\begin{eqnarray}
\nonumber
\Omega_{1,\alpha\beta}&=&\partial_{\alpha}w_{\beta}+\partial_{\beta}w_{\alpha}
-\delta_{\alpha\beta}\partial_{\gamma}w_{\gamma} \\
\nonumber
\Omega_{2,\alpha\beta}&=&2\partial_{\alpha}\partial_{\beta}\rho
-\delta_{\alpha\beta}\partial^2_{\gamma}\rho \\
\nonumber
\Omega_{3,\alpha\beta}&=&2w_{\alpha}w_{\beta}
-\delta_{\alpha\beta}w^2 \\ 
\nonumber
\Omega_{4,\alpha\beta}&=&w_{\alpha}\partial_{\beta}\rho+w_{\beta}\partial_{\alpha}\rho
-\delta_{\alpha\beta}w_{\gamma}\partial_{\gamma}\rho \\
\label{OMEGA_DEF}
\Omega_{5,\alpha\beta}&=&2(\partial_{\alpha}\rho)(\partial_{\beta}\rho)
-\delta_{\alpha\beta}(\partial_{\gamma})^2\,.
\end{eqnarray}
The tensor $\Omega_1$ is the viscous stress tensor of a two-dimensional fluid.
The transport coefficients in Eq.\ (\ref{TENSOR1}) are given in 
Table \ref{TAB2}. They depend on the following variables,
\begin{eqnarray}
\nonumber
p&=&
{4\over \eta} \sin{(\eta)}\sum_{n=1}^N
{{\rm e}^{-M_R}\over n!}n^2
M_R^{n-1} J_1(n) \\ 
\nonumber
q&=&
{4\pi \gamma^2\over \eta} \sin{(\eta)}\sum_{n=2}^N
{{\rm e}^{-M_R}\over n!}n^2(n-1)
M_R^{n-2} J_2(n)\\ 
\label{DEF_PQ}
S&=&
{8\pi\gamma^2\over \eta }\sin{\eta\over 2}
\sum_{n=2}^N
{{\rm e}^{-M_R}\over n!}
n^2(n-1) M_R^{n-2} J_3(n) \\
\nonumber
\Gamma&=&
{8\pi^2\gamma^4\over 3 \eta}\sin{\eta\over 2}
\sum_{n=3}^N
{{\rm e}^{-M_R}\over n!}
n^2 (n-1)(n-2)M_R^{n-3} J_4(n)
\end{eqnarray}
\begin{table}
\begin{center}
{\renewcommand{\arraystretch}{1.4}
\large
\begin{tabular}{|r||c|c|c|}
\hline
$j$ & $h_j$      &    $q_j$      &      $k_j$     \\
\hline
\hline
$1$ & ${1+p\over 8(p-1)}$          &  ${S\over 2(p-1)}$     &  $ {S\over 8 (p-1)}$ \\ 
\hline
$2$ & $-{p^2+10p+1\over 96(p-1)^2}$& $-{S\over 4(p-1)^2}$   & $-{S (p+5)\over 96(p-1)^2}$ \\
\hline
$3$ & $-{q\over 2(p-1)} $          & $\Gamma-{Sq\over p-1}$ & ${\Gamma\over 4}-{Sq\over 4(p-1)}$ \\
\hline  
$4$ & ${q(1+p)\over 4(p-1)^2}$     & ${\Gamma\over 2}-{Sq(p-3)\over 2(p-1)^2}$ & ${\Gamma\over 12}-{Sq(p-4)\over 12(p-1)^2}$ \\
\hline
$5$ & $-{q(p^2+10p+1)\over 48(p-1)^3}$ & ${\Gamma\over 24}-{Sq(p^2-2p+13)\over24(p-1)^3}$ & $-{Sq(p+5)\over 48(p-1)^3}$ \\
\hline
\end{tabular}
}
\caption{The transport coefficients $h_j$, $q_j$ and $k_j$,
defined in Eq. (\ref{TENSOR1}), are expressed as functions of 
$\Gamma$, $S$, $p$, $q$, see Eq. (\ref{DEF_PQ}).
}
\label{TAB2}
\end{center}
\vspace*{-5ex}
\end{table}
where $\gamma$ is the ratio of the interaction radius to the mfp, $\gamma=R/\tau v_0$.
Eq.\ (\ref{FINAL_EQ}) is consistent with the one 
postulated in Ref.\ \cite{toner_95_98} but
contains 
additional gradient terms. 
It has a homogeneous flocking solution: $\vec{w}=w_0\,{\bf \hat{n}}$ and $\rho=\rho_0$. The amplitude of the flow is given by $w_0=\sqrt{(1-\lambda)/q_3}$. 
In order to study the spontaneous onset of collective motion,
a perturbation around this 
state
is considered,
$\rho({\bf x},t)=\rho_0+\delta\rho\,{\rm e}^{i{\bf k}\cdot {\bf x}+\omega t}$,
$\vec{w}({\bf x},t)=w_0\,{\bf \hat{n}}+\delta\vec{w}\,{\rm e}^{i{\bf k}\cdot {\bf x}+\omega t}
$,
and Eq.\ (\ref{FINAL_EQ}) is linearized in $\delta \rho$ and $\delta\vec{w}$.
The characteristic equation for
the growth rate $\omega(\vec{k})$ 
describes three possible modes. 
I found that in a small window, $\eta_S<\eta<\eta_C$, directly below the onset of flocking, one of the longitudinal modes is always unstable against long wavelength
perturbations: the real part of $\omega$ is positive for $0<k<k_0$ as shown in Fig.\ 1b).
A similar instability was reported by Bertin {\em et al.} \cite{bertin_06}.
Chat{\'e} {\em et al.} \cite{chate_04_08}
found numerically
that the order/disorder transition is discontinuous for
system sizes $L$ larger than 
the crossover length $L^{*}$.
Assuming that the long wave instability is the reason for this finite size effect, 
I calculated the largest value of $k_0$ within the narrow instability window
at constant density, $k^{*}$,
in order to obtain a lower bound for  
$L^{*}$.
Plotting $2\pi/k^{*}$ gives the lower curve in the insert of Fig. 1b).
An upper bound was obtained by determining the wave number $k_{max}$ where 
the growth rate has the largest value inside the instability window. 
The upper curve in the insert shows
$2\pi/k_{max}$ as a function of density.
The minimum around $M\approx 2$ and the divergences at small and large densities
are consistent with numerical results 
\cite{chate_04_08}.

To see what happens to a growing 
perturbation beyond the 
linear instability, the continuity equation and Eq.\ (\ref{FINAL_EQ}) 
were integrated
on a $L\times L$ lattice with periodic boundaries by means of a predictor-corrector scheme 
\cite{mccormack_82}.
These simulations confirmed  
that the ordered phase is stable for small system sizes $L<2\pi/k_0$. 
For slightly larger system sizes one observes a stable, inhomogeneous
steady state with a global order parameter, $\langle \vec{w}\rangle=\int\vec{w}\,d{\bf x}/L^2$, larger than the amplitude of the homogeneous state, $w_0$.
Finally, for much
larger system sizes, it turns out that 
the system is both linearly and nonlinearly unstable 
for $\eta_S<\eta<\eta_C$.
Longitudinal perturbations grow without bound; they do not lead
to stable solitons as suggested 
in Ref. \cite{bertin_06}.
However, direct simulations of the VM at large mfp do show solitary structures
such as traveling high-density bands in a window just below the transition \cite{chate_04_08,ihle_10}.
At lower noise 
these structures disappear.
Identifying this ``solitary'' window with the instability window,
its size can be predicted by the current theory which takes 
all the details of the VM 
such as multi-body interactions
into account.
However, inside this window, the hydrodynamic equations
are driven out of the range of their validity and 
are not suited to describe solitons. 
Nagy et al \cite{vicsek_95_07} did not see high-density bands at small velocity $v_0$.
To treat this limit of small mfp theoretically, 
one has to abandon the molecular chaos approximation i.e. 
go beyond the mean-field approximation, which is outside the scope
of this paper.

In summary, a first-principle derivation of the hydrodynamic equations
of the VM by means of a novel kinetic theory 
is presented
and a stability analysis of the resulting equations, Eq. (\ref{FINAL_EQ}),
is performed.
The mean-field phase diagram for arbitrary density is calculated.
It agrees within a few percent with simulation results and is 
shown to be independent 
of the particle speed in the large mfp limit.
It is also shown that the continuous theory of \cite{bertin_06} fails to
reproduce the phase diagram of the VM and that one has to
explicitely incorporate the discrete time dynamics and 
genuine multi-body interactions in order to achieve agreement. 
The theory presented here is consistent with numerical studies \cite{vicsek_95_07,chate_04_08}, 
and suggests
the following picture 
of the nature of the flocking transition in the {\it large mfp limit} considered here:
At $\eta=\eta_C$ a homogeneous ordered state bifurcates continuously from
the disordered state.  
At the threshold, this state is unstable to longitudinal, 
long wavelength fluctuations. 
Perturbations from a large range of wave numbers $k<k_0$ 
become
unstable, already in close vicinity to the threshold. 
The transition appears to be continuous in small systems 
but 
becomes a discontinuos transition in large systems due to the emergence
of density waves which abruptly increase the global order parameter.
An estimate of the system size $L^{*}$, above which
the discontinuous nature of the transition is expected to emerge, is given.
This length is found to diverge
at small and large densities, consistent with 
numerical results. 

Support
from the National Science Foundation under grant No.
DMR-0706017 
is gratefully acknowledged.
I thank F. J{\"u}licher, L.S. Schulman, H. Chat{\'e},
A. Denton, A. Wagner and D. Kroll
for valuable discussions.

\end{document}